# From symmetry break to Poisson point process in 2D Voronoi tessellations: the generic nature of hexagons

Valerio Lucarini

ADGB - Department of Physics
University of Bologna
Viale Berti Pichat 6/2
40127, Bologna, Italy

CINFAI
Via Viviano Venanzi 15
62032, Camerino, Italy

Email: lucarini@adgb.df.unibo.it

## Abstract

We bridge the properties of the regular square and honeycomb Voronoi tessellations of the plane to those of the Poisson-Voronoi case, thus analyzing in a common framework symmetry-break processes and the approach to uniformly random distributions of tessellation-generating points. We resort to ensemble simulations of tessellations generated by points whose regular positions is perturbed through a Gaussian noise whose adimensional strength is controlled by the parameter $\alpha$. We analyze the number of sides, the area, and the perimeter of the Voronoi cells. For $\alpha>0$, hexagons constitute the most common class of cells, and 2-parameter gamma distributions provide an efficient description of statistical properties of the analyzed geometrical characteristics. The symmetry break induced by the introduction of noise destroys the square tessellation, which is structurally unstable, whereas the honeycomb hexagonal tessellation is very stable and all Voronoi cells are hexagon for small but finite noise with $\alpha<0.1$. Several statistical signatures of the symmetry break are evidenced. Already for a moderate amount of Gaussian noise ($\alpha>0.5$), memory of the specific initial unperturbed state is lost, because the statistical properties of the two perturbed regular tessellations are indistinguishable. When $\alpha>2$, results converge to those of Poisson-Voronoi tessellations. The geometrical properties of $n$-sided cells change with $\alpha$ until the Poisson-Voronoi limit is reached for $\alpha>2$; in this limit the Desch law for perimeters is confirmed to be not valid and a square root dependence on $n$, which allows an easy link to the Lewis law for areas, is established. Finally, the ensemble mean of the cells area and perimeter restricted to the hexagonal cells coincides with the full ensemble mean; this might imply that the number of sides acts as a *thermodynamic state variable* fluctuating about $n=6$, and this reinforces the idea that hexagons, beyond their ubiquitous numerical prominence, can be taken as *generic* polygons in 2D Voronoi tessellations.

# Introduction

Given a discrete set of points $X$ in an Euclidean N-dimensional space, we have that for almost any point $a$ of such a space there is one specific point $x \in X$ which is closest to $a$. Some point $x$ may be equally distant from two or more points of $X$. If $X$ contains only two points, $x_1$ and $x_2$, then the set of all points with the same distance from $x_1$ and $x_2$ is a hyperplane, which has codimension 1. The hyperplane bisects perpendicularly the segment from $x_1$ and $x_2$. In general, the set of all points closer to a point $x_i \in X$ than to any other point $x_j \neq x_i$, $x_j \in X$ is the interior of a convex (N-1)-polytope usually called the Voronoi cell for $x_i$. The set of the (N-1)-polytopes $\Pi_i$, each corresponding to, and containing, one point $x_i \in X$, is the Voronoi tessellation corresponding to $X$, and provides a partitioning of the considered N-dimensional space (Voronoi 1907, 1908). As well known, the Delaunay triangulation (Delaunay 1934) gives the dual graph of the Voronoi tessellation (Okabe et al. 2000). Voronoi cells can be defined also for non-Euclidean metric spaces, but in the general case the existence of a Voronoi tessellation is not guaranteed.

Since the Voronoi tessellation creates a one-to-one optimal - in the sense of minimum distance - correspondence between a point and a polytope, 2D and 3D Voronoi tessellations have been considered for a long time for applications in several research areas, such as telecommunications (Sortais et al., 2007), biology (Finney 1975), astronomy (Icke 1996), forestry (Barrett 1997) atomic physics (Goede et al. 1997), metallurgy (Weaire et al., 1986), polymer science (Dotera 1999), materials science (Bennett et al., 1986). In solid-state physics, the Voronoi cells of the single component of a crystal are known as Wigner-Seitz unit cells (Ashcroft and Mermin 1976). In a geophysical context, Voronoi tessellations have been widely used to analyze spatially distributed observational or model output data (Tsai et al. 2004); in particular, they are a formidable tool for performing arbitrary space integration of sparse data, without adopting the typical procedure of adding spurious information, as in the case of linear or splines interpolations, etc. (Lucarini et al. 2007). Actually, in this regard that Thiessen and Alter, with the purpose of computing river basin water balances from irregular and sparse rain observations, discovered independently for the 2D case the tessellation introduced by Voronoi just few years earlier (Thiessen and Alter 1911). Moreover, recently a connection has been established between the Rayleigh-Bènard convective cells and Voronoi cells, with the hot spots (strongest upward motion of hot fluid) of the former basically coinciding with the points *generating* the Voronoi cells, and the locations of downward motion of cooled fluid coinciding with the sides of the Voronoi cells (Rapaport 2006).

The quest for achieving low computational cost for actually evaluating the Voronoi tessellation of a given discrete set of points $X$ is ongoing and involves an extensive research performed within various scientific communities (Bowyer 1981; Watson 1981; Tanemura et al. 1983; Barber et al. 1996; Han and Bray 2006). The theoretical investigation of the statistical properties of general N-dimensional Voronoi tessellations, which has a great importance in applications, has proved to be a rather hard task, so that direct numerical simulation is the most extensively adopted investigative approach. For a review of the theory and applications of Voronoi tessellations, see Aurenhammer (1991) and Okabe et al. (2000).

A great deal of theoretical and computational work has focused on a more specific and tractable problem, that of studying the statistical properties of the geometric characteristics of Poisson-Voronoi tessellations. These are Voronoi tessellations obtained for a random set of points $X$ generated as output of a homogeneous Poisson point process. This problem has a great relevance at practical level because it corresponds, e.g., to studying crystal aggregates with random nucleation sites and uniform growth rates. Exact results concerning the mean statistical properties of the interface area, inner area, number of vertices, etc. of the Voronoi cells have been obtained for 2+ dimensional Euclidean spaces (Meijering 1953; Christ et al., 1982; Drouffe and Itzykson 1984; Finch 2003; Calka 2003). Recently, some important results have been obtained for the 2D case (Hilhorst 2006). Several computational studies, performed considering a quite wide range of number of 2D and 3D Voronoi cells, have found results that basically agree with the theoretical findings, and, moreover, have shown that both 2-parameter (Kumar et al. 1992) and 3-parameter (Hinde and Miles 1980) gamma distributions fit up to a high degree of accuracy the empirical pdfs of the number of vertices, of the perimeter and of the area of the cells. Very extensive and more recent calculations have basically confirmed these results (Tanemura 2003). In spite of several attempts, the ab-initio derivation of the pdf of the geometrical properties of Poisson-Voronoi tessellations have not been yet obtained, except in asymptotic regimes (Hilhorst 2005). In the last years, various studies have focused on the geometrical properties of Voronoi tessellations resulting from non strictly Poissonian random processes. In particular, given the obvious applicative value (as for packaging problems), a great deal of emphasis has been put on tessellations resulting from points which are randomly distributed in the space but which also cannot be closer than a given distance $\delta$ - a sort of hard-core nuclei hypothesis (Zhu et al. 2001; Senthil Kumar and Kumaran 2005). Whereas the $\delta=0$ corresponds to the Poisson-Voronoi case, it is observed that by increasing $\delta$ the *degree of randomness* of the tessellation is decreased - the pdfs of the statistical properties of the geometrical characteristics become more and more peaked - until at a certain critical value of $\delta$ a regular tessellation, which in the 2D case is the hexagonal honeycomb, is obtained. In any case, it is found that the gamma distributions provide excellent fits for a very large range of values of δ (Zhu et al. 2001).

In this paper we want to explore a somewhat different problem of parametric dependence of the Voronoi tessellation statistics. We start from two regular polygonal tessellations of the plane, the honeycomb hexagonal tessellation and the square tessellations. They are obtained by setting the points $x_i$ as vertices of regular triangles and squares, respectively. Using an ensemble-based approach, we study the break-up of the symmetry of the two systems and quantitatively evaluate how the statistical properties of the geometrical characteristics of the resulting 2D Voronoi cells change when we perturb with a space-homogeneous Gaussian noise of increasing intensity the positions of the points $x_i$. Our paper is organized as follows. In section 2 we describe the methodology of work and the set of numerical experiments performed, In section 3 we show our results. In section 4 we present our conclusions and perspectives for future work.

## Data and Methods

### *Scaling Properties of Voronoi tessellations*

Let's first consider a homogeneous Poisson point process $\Psi$ generating as output a random set of points $X$ in such a way that the expectation value for the number of points $x_i$ belonging, without loss of generality, to a square region $\Gamma$ is $\rho_0|\Gamma| = \rho_0 R^2$, where $\rho_0$ is the intensity of the process, $|\Gamma|$ is the measure of $\Gamma$ and $R$ is the side of the square, whereas the fluctuations are of the order of $\sqrt{\rho_0|\Gamma|} = \sqrt{\rho_0} R$. If $\rho_0|\Gamma| \gg 1$, we are in the thermodynamic limit and the number of Voronoi cells $C_i$ inside $\Gamma$ is $N_V \approx \rho_0|\Gamma|$; in other terms the contributions due to the cells crossing the boundary of $\Gamma$ are negligible. Keeping $\rho_0$ fixed and considering larger values of $R$, the approximation further improves, and similarly occurs when keeping $R$ fixed and increasing $\rho_0$. In this limit, theoretical results suggest that the expectation value – where the statistics is computed over the $N_V$ cells and over an ensemble of random processes – of number of sides of the Voronoi cells inside $\Gamma$ is $\langle \mu(n)|_V \rangle = 6$, of the area of the Voronoi cell is $\langle \mu(A)|_V \rangle = 1/\rho_0$, and of the perimeter of a Voronoi cell $\langle \mu(P)|_V \rangle = 4/\sqrt{\rho_0}$. For clarity's sake, we specify that the expression $\mu(Y)$ ($\sigma(Y)$) refers to the mean value (standard deviation) of the variable $Y$ over the $N_V$ cells for the single realization of the random process, the expression $\langle E \rangle$ ($\delta[E]$), instead, indicates the ensemble mean (standard deviation) of the random variable $E$. Therefore, the ensemble mean statistical properties of the

Poisson-Voronoi tessellation are intensive, in the sense that they depend only on the average density of points (and basically of Voronoi cells) and not on the shape (in spite of our initial assumption) or size of $\Gamma$, once a sufficiently large number of points is considered. Moreover, we also have that the ensemble mean of the standard deviation scales with respect to $\rho_0$ as the mean value, so that for each quantity $Y$ we have that $\langle\sigma(Y)\rangle/\langle\mu(Y)\rangle$ does not depend on $\rho_0$. It is then clear that in this case it is possible to restrict the analysis, *e.g.*, to the square $\Gamma_1 = [0,1]\times[0,1]$.

If the random point process $\Phi$ generates a periodic distribution of the points $x_i$ in the plane with discrete translational symmetry generated by the lattice vectors $\vec{v}_1$ and $\vec{v}_2$, we expect that, if $R \gg |\vec{v}_1|,|\vec{v}_2|$, the expectation value of the number of $x_i$'s belonging to a square region $\Gamma$ can also be expressed as $\rho_0|\Gamma| = \rho_0 R^2$, independently of the position of $\Gamma$ in the plane, where $\rho_0$ is the coarse-grained intensity of the process. Considering that the geometrical procedure leading to the construction of the Voronoi tessellation is a local and a self-similar one, and no *long-distance interactions* are considered. the ensemble mean statistical properties of the Voronoi tessellation do not depend on the size and position of $\Gamma$. Also in this more general case, if $\rho_0 \gg 1$ and $1 \gg |\vec{v}_1|,|\vec{v}_2|$, it is then possible to restrict the analysis, *e.g.*, to the square $\Gamma_1 = [0,1]\times[0,1]$ in the Cartesian plane. If, after having generated the Voronoi tessellation, we perform a linear $\lambda$-rescaling of the coordinates such that $\Gamma_1 = [0,1]\times[0,1]$ transforms into $\Gamma_\lambda = [0,\lambda]\times[0,\lambda]$, we have that the measure of any 1D (2D) object grows by factor of $\lambda$ ($\lambda^2$), and $\rho_0 \to \rho_0/\lambda^2$, whereas the number of sides of each cell is not altered. This procedure is equivalent, when ensemble means are considered, to refrain from any rescaling of the coordinates and dividing instead $\rho_0$ by the factor $\lambda^2$, because the size of the region $\Gamma$ is not relevant. Therefore, we deduce that in the thermodynamics regime for all random processes $\Phi$ with the above mentioned characteristics, $\langle\mu(n)\rangle,\langle\sigma(n)\rangle \propto (\rho_0)^0$, $\langle\mu(A)\rangle,\langle\sigma(n)\rangle \propto 1/\rho_0$, and $\langle\mu(P)\rangle,\langle\sigma(P)\rangle \propto 1/\sqrt{\rho_0}$. Therefore, by multiplying the ensemble mean estimators of the mean and standard deviation of the area (perimeter) of the Voronoi cells times $\rho_0$ ($\sqrt{\rho_0}$), we obtain universal functions.

## *Simulations*

As a starting point, we consider two regular tessellations of the plane. If we consider a regular square gridding of the points $x_i$ with sides $l = |\vec{v}_1| = |\vec{v}_2|, \vec{v}_1 \perp \vec{v}_2$, the Voronoi cell $\Pi_i$ corresponding to $x_i$ is given by the square centered in $x_i$ with the same side length and orientation as the $x_i$ grid. If

$l_Q = 1/\sqrt{\rho_0} = |\vec{v}_1| = |\vec{v}_2|$, we will have $\rho_0$ points – and $\rho_0$ corresponding square Voronoi cells - in $\Gamma_1 = [0,1] \times [0,1]$. Similarly, a regular hexagonal honeycomb tessellation featuring $\rho_0$ points and approximately $\rho_0$ corresponding square Voronoi cells in $\Gamma_1 = [0,1] \times [0,1]$ is obtained by using a gridding of points set as regular triangles with sides $l_H = \sqrt{2/\rho_0 \sqrt{3}} = |\vec{v}_1| = |\vec{v}_2|$.

For each of the two regular griddings, we then introduce a symmetry-breaking 2D-homogeneous ε-Gaussian noise, which randomizes the position of each of the points $x_i$ about its deterministic position with a spatial variance $|\varepsilon^2|$. We express $|\varepsilon^2| = \alpha^2 l_Q^2 = \alpha^2/\rho_0$, thus expressing the mean squared displacement as a fraction $\alpha^2$ of the inverse of the density of points, which is the natural squared length scale. Note that in all cases, when ensembles are considered, the distribution of the $x_i$ is still periodic.

For each of the two regular griddings, we then perform our statistical analyses by considering M = 1000 members of the ensemble of Voronoi tessellations generated for each value of α ranging from 0 to 5 with step 0.01. The actual simulations are performed by using, within a customized routine, the MATLAB7.0® function voronoin.m, which implements the algorithm introduced by Barber et al. (1996), to a set of points $x_i$ having density $\rho_0 = 10000$. Tessellation has been performed starting from points $x_i$ belonging to the square $[-0.2, 1.2] \times [-0.2, 1.2] \supset \Gamma_1 = [0,1] \times [0,1]$, but only the cells belonging to $\Gamma_1$ have been considered for evaluating the statistical properties, in order to basically avoid $\rho_0$ depletion in the case of large values of α due to one-step Brownian diffusion of the points nearby the boundaries.

By definition, if α = 0 we are in the deterministic case. We study how the statistical properties of *n*, *P*, and *A* of the Voronoi cells change with α, covering the whole range going from the symmetry break, occurring when α becomes positive, up to the progressively more and more uniform distribution of $x_i$, obtained when α is large with respect to 1 and the distributions of nearby points $x_i$ overlap more and more significantly. The distributions of *n*, *P*, and *A* are fitted using a 2-parameter gamma distribution with the MATLAB7.0® function gammafit.m, which implements a maximum likelihood method.

## Results

We expect that the exploration of the parametric range from $\alpha = 0$ to $\alpha = 5$ should allow us to join on the two extreme situations of perfectly deterministic, regular tessellation, to the tessellation resulting from a set of points *X* generated with a Poisson point process.

In the deterministic $\alpha = 0$ case, it is easy to deduce the properties of the Voronoi cells from basic Euclidean geometry. For square tessellation, we have $\langle \mu(n)|_{\alpha=0} \rangle = \mu(n)|_{\alpha=0} = 4$, $\langle \mu(P)|_{\alpha=0} \rangle = \mu(P)|_{\alpha=0} = 4 \times 10^{-2} = 4/\sqrt{\rho_0}$, and $\langle \mu(A)|_{\alpha=0} \rangle = \mu(A)|_{\alpha=0} = 4 \times 10^{-4} = 1/\rho_0$, where the $\alpha$-dependence of the statistical properties is indicated. For honeycomb tessellation, we have $\langle \mu(n)|_{\alpha=0} \rangle = \mu(n)|_{\alpha=0} = 6$, $\langle \mu(P)|_{\alpha=0} \rangle = \mu(P)|_{\alpha=0} = \sqrt{24/\sqrt{3}} \times 10^{-2} = \sqrt{24/(\sqrt{3}\rho_0)}$, and $\langle \mu(A)|_{\alpha=0} \rangle = \mu(A)|_{\alpha=0} = 4 \times 10^{-4} = 1/\rho_0$. Of course, in both cases, given the regular pattern in space, all cells are alike, and given the deterministic nature of the tessellation, there are no fluctuations within the ensemble.

## *Number of sides of the cells*

In the case of the regular square tessellation, the introduction of a minimal amount of symmetry-breaking noise acts as singular perturbation for the statistics of $\mu(n)|_\alpha$ and $\sigma(n)|_\alpha$, since $\langle \mu(n)|_\alpha \rangle$ and $\langle \sigma(n)|_\alpha \rangle$ are discontinuous in $\alpha = 0$. We have that $\langle \mu(n)|_{\alpha=0} \rangle = 4 \neq 6 = \langle \mu(n)|_{\alpha=0^+} \rangle$ and $\langle \sigma(n)|_{\alpha=0} \rangle = 0 \neq 0.93 \approx \langle \sigma(n)|_{\alpha=0^+} \rangle$, where with $\alpha = 0^+$ we indicate the right limit to 0 with respect to the parameter α. Similarly, the ensemble fluctuations $\delta[\mu(n)|_\alpha]$ and $\delta[\sigma(n)|_\alpha]$ are discontinuous functions in $\alpha = 0$, since they reach a finite value > 0 as soon as the noise is switched on. This proves that such a tessellation is structurally unstable. Note that the simulations have been performed considering a very high resolution on the parameter α for $\alpha \approx 0$. Considering larger values of α, we have that $\langle \sigma(n)|_\alpha \rangle$ is basically constant up to $\alpha \approx 0.35$, where its value begins to quickly increase before reaching the asymptotic value $\langle \sigma(n)|_\alpha \rangle \approx 1.33$ for $\alpha > 2$, which essentially coincides with what obtained in the Poisson-Voronoi case. The function $\langle \mu(n)|_\alpha \rangle$ is, instead, remarkably constant within few permils around the value of 6 for $\alpha > 0$, which shows that this value is not specific to the Poisson-Voronoi case, but rather depending upon the topology of the plane. In Fig. 1 we plot the functions $\langle \mu(n)|_\alpha \rangle$ and $\langle \sigma(n)|_\alpha \rangle$, whereas the half-width of the error bars are twice the corresponding values of $\delta[\mu(n)|_\alpha]$ and $\delta[\sigma(n)|_\alpha]$. The Poisson-Voronoi values are indicated for reference.

Except for the singular case $\alpha = 0$, for all $\alpha > 0$ the distribution of the number of sides of the cells obey up to a very high degree of precision a 2-parameter gamma distribution:

$$f(x;k,\theta) = N_V x^{k-1} \frac{\exp[-x/\theta]}{\theta^k \Gamma(k)} \qquad (1),$$

where $\Gamma(k)$ is the usual gamma function and $N_V \approx \rho_0$ is, by definition, the normalization factor. We have, in the case of unbiased estimators (as in this case), that $\mu(n)|_\alpha = k(n)|_\alpha \theta(n)|_\alpha$ and $\sigma(n)|_\alpha = \sqrt{k(n)|_\alpha \theta^2(n)|_\alpha}$. We observe that both functions $\langle k(n)|_\alpha \rangle$ and $\langle \theta(n)|_\alpha \rangle$ (not shown) are basically constant for $0 < \alpha < 0.35$, then for larger values of α $\langle k(n)|_\alpha \rangle$ increases and $\langle \theta(n)|_\alpha \rangle$ decreases in such a way that their product is constant, because $\langle \mu(n)|_\alpha \rangle = \langle k(n)|_\alpha \theta(n)|_\alpha \rangle \approx \langle k(n)|_\alpha \rangle \langle \theta(n)|_\alpha \rangle$, and for $\alpha > 2$ the two functions become closer and closer to their asymptotic values, which agrees remarkably well with what obtained for the Poisson-Voronoi case. These results suggest that, topologically speaking, the *route to randomness* from the square regular tessellation to the Poisson-Voronoi case goes through a transition involving a *stable* – with respect to the complete statistics of the number of sides - *pattern of cells*, which persists for the finite range $0 < \alpha < 0.35$. In this range, hexagons dominate and their fraction is constant, whereas for larger values of $\alpha$, the fraction of hexagon declines but is still dominant.

When considering the regular hexagon honeycomb tessellation, the impact of introducing noise in the position of the points $x_i$ is quite different from the previous case. Results are also shown in Fig. 1. The first observation is that an infinitesimal noise does not effect at all the tessellation, in the sense that all cells remain hexagons. Moreover, even finite-size noise basically does not distort cells in such a way that figures other than hexagons are created. We have not observed non *n*=6 cells for up to $\alpha \approx 0.12$ in any member of the ensemble. This has been confirmed also considering larger densities (e.g. $\rho_0 = 1000000$). It is more precise, though, to frame the structural stability of the hexagon tessellation in probabilistic terms: the creation of a non-hexagons is very unlikely for the considered range. Since the Gaussian noise induces for each point $x_i$ a distribution with – an unrealistic- non-compact support, in principle it is possible to have outliers that, at local level, can distort heavily the tessellation.

For $\alpha > 0.12$, $\langle \sigma(n)|_\alpha \rangle$ is positive and increases monotonically with $\alpha$; this implies that the fraction of hexagons decreases monotonically with $\alpha$. For $\alpha > 0.5$ the value of $\langle \sigma(n)|_\alpha \rangle$ is not distinguishable from that obtained in the previous case of perturbed square tessellation. Similarly to the previous case, for all values of $\alpha$ we have that $\langle \mu(n)|_\alpha \rangle \approx 6$ within few permils; such constraint is confirmed to be very strong and quite general. For $\alpha > 0.12$, the empirical distribution of the number of sides of the cells can be modelled quite efficiently with 2-parameter gamma distributions. We have that $\langle k(n)|_\alpha \rangle$ decreases with $\alpha$, whereas the converse is true for $\langle \theta(n)|_\alpha \rangle$; obviously, the estimates of the two parameters are in statistical agreement with what obtained in the square tessellation for $\alpha > 0.5$. This implies that from a statistical point of view, the variable number of edges loses memory of its unperturbed state already for a rather low amount of Gaussian noise, well before becoming undistinguishable from the fully random Poisson case.

## *Area and Perimeter of the cells*

For both the perturbed square and honeycomb hexagonal tessellation, the parametric dependence on $\alpha$ of the statistical properties of the area of the Voronoi cells is more regular than for the case of the number of sides. Results are shown in Fig. 2.

In general, the ensemble mean value $\langle \mu(A)|_\alpha \rangle$ of the area of the Voronoi cells is, basically by definition, constrained to be $\langle \mu(A)|_\alpha \rangle = 1/\rho_0 = 10^{-4}$ for all values of $\alpha$, and we observe that for both perturbed tessellation its fluctuations $\delta[\mu(A)|_\alpha]$ have, for $\alpha > 0$, a constant value, coinciding with that observed in the Poisson-Voronoi case. The $\alpha$-dependence of $\langle \sigma(A)|_\alpha \rangle$ is more interesting. We first note that the two functions $\langle \sigma(A)|_\alpha \rangle$ computed from the two perturbed tessellation are basically coincident, and the same occurs for $\delta[\sigma(A)|_\alpha]$. This implies that the impact of adding noise in the system in the variability of the area of the cells is quite general and does not depend on the unperturbed patters. We can be confident of the generality of this result also because for relatively small values of $\alpha$ (say, $\alpha < 0.5$), $\langle \sigma(A)|_\alpha \rangle$ has a specific functional form reminding of symmetry breaking behaviour: in such a range we have that $\langle \sigma(A)|_\alpha \rangle \approx \langle \sigma(A)|_V \rangle \times \alpha$. For $\alpha > 2$, $\langle \sigma(A)|_\alpha \rangle$ is almost indistinguishable from the Poisson-Voronoi value, so that we can estimate an asymptotic value $\langle \sigma(A)|_V \rangle \approx 0.53/\rho_0 = 5.3 \times 10^{-5}$.

Results for the statistical estimators of the perimeter of the Voronoi cells are shown in figure 3. When considering the perturbed square tessellation, $\langle \mu(P)|_\alpha \rangle$ basically coincides with that of the Poisson-Voronoi case for $\alpha > 1$. Note that also $\langle \mu(P)|_{\alpha=0} \rangle = 4/\sqrt{\rho_0} = 4 \times 10^{-2} = \langle \mu(P)|_V \rangle$, but anyway $\langle \mu(P)|_\alpha \rangle$ is a function with some interesting structure: for $\alpha = \alpha_m \approx 0.25$ $\langle \mu(P)|_\alpha \rangle$ features a distinct minimum $\langle \mu(P)|_{\alpha=\alpha_m} \rangle \approx 0.975 \langle \mu(P)|_V \rangle$, whereas for $\alpha = \alpha_M \approx 0.75$ a maximum for $\langle\mu(P)|_\alpha\rangle$ is realized, with $\langle\mu(P)|_{\alpha=\alpha_M}\rangle\approx 1.01\langle\mu(P)|_V\rangle$. The unperturbed honeycomb hexagonal tessellation is optimal in the sense of perimeter-to-area ratio, and, when noise is added the corresponding function $\langle\mu(P)|_\alpha\rangle$ increases quadratically (not shown) with $\alpha$ for $\alpha < 0.3$, whereas for $\alpha > 0.5$ its value coincides with what obtained starting from the regular square tessellation. We deduce that there is, counter-intuitively, a specific amount of noise (for $\alpha = \alpha_m$) which optimizes the mean perimeter-to-area ratio for the regular square tessellation, whereas, for $\alpha = \alpha_M$ the opposite is realized for both tessellation. When considering the functions $\langle\sigma(P)|_\alpha\rangle$, we are in a similar situation as for the statistics of mean cells area: the result of the impact of noise is the same for both tessellations, and for $\alpha < 0.5$, $\langle\sigma(P)|_\alpha\rangle$ is proportional to $\alpha$, with $\langle\sigma(P)|_\alpha\rangle \approx \langle\sigma(P)|_V\rangle \times \alpha$. Moreover, for $\alpha > 2$, $\langle\sigma(P)|_\alpha\rangle$ becomes undistinguishable from the asymptotic value realized for Poisson-Voronoi process $\langle\sigma(A)|_V\rangle \approx 0.98/\sqrt{\rho_0} = 9.8\times 10^{-3}$.

For $\alpha > 0$, the empirical pdfs of cells area and perimeter can be fitted very efficiently using 2-parameter gamma distributions. The wide-extent effectiveness of using 2-parameter gamma distributions for fitting the statistics of all of the geometric properties of 2D Voronoi cells, noted by Zhu et al. (2001) for another sort of parametric investigation, is confirmed also in this case. The behaviour of $\langle\mu(k)|_\alpha\rangle$ and $\langle\mu(\theta)|_\alpha\rangle$ for small values of $\alpha$ can be obtained as follows. Since $\langle\mu(Y)|_\alpha\rangle = \langle k(Y)|_\alpha \theta(Y)|_\alpha\rangle \approx \langle k(Y)|_\alpha\rangle\langle\theta(Y)|_\alpha\rangle$ (with $Y = P, A$) is basically constant (within few percents for both tessellations), and $\langle\sigma^2(Y)|_\alpha\rangle \approx \langle k(Y)|_\alpha \theta^2(Y)|_\alpha\rangle \approx \langle k(Y)|_\alpha\rangle\langle\theta(Y)|_\alpha\rangle^2$ with $\langle\sigma^2(Y)|_\alpha\rangle \approx \langle\sigma(Y)|_\alpha\rangle^2 \propto \alpha^2$, we have that $\langle k(Y)|_\alpha\rangle \propto \alpha^{-2}$ and $\langle\theta(Y)|_\alpha\rangle \propto \alpha^2$.

## Area and perimeter of n-sided cells

A subject of intense investigation has been the characterization of the geometrical properties of n-sided cells; see Hilhorst (2006) and references therein for a detailed discussion. We have then computed the for the considered range of $\alpha$ the quantities $\langle\mu(A)|_\alpha\rangle_n$, $\delta[\mu(A)|_\alpha]_n$, $\langle\mu(P)|_\alpha\rangle_n$, and $\delta[\mu(P)|_\alpha]_n$, obtained by stratifying the outputs of the ensemble of simulations with respect to the number of sides $n$ of the resulting cells. The 2-standard deviation confidence interval centered around the ensemble mean is shown as a function of n in Fig. 4 for the area and the perimeter of the cells, for selected values of $\alpha$. Note that for larger values of $n$ the error bar is larger because the number of occurrences of $n$-sided cells is small.

The results of the two perturbed regular tessellations basically agree for $\alpha > 0.5$, thus confirming what shown previously. This implies that already a moderate amount of noise provides an efficient mixing, which allows the convergence of the statistical properties of tessellation resulting from rather different regular, unperturbed *parent tessellation*s, with very different $\alpha \approx 0$ behaviour.

In particular, for $\alpha > 2$, the results coincide with what resulting from the Poisson-Voronoi case. Firstly, we verify the Lewis law, *i.e.* $\langle \mu(A)|_\alpha \rangle_n \approx a_1/\rho_0 (n + a_2)$. Our data give $a_1 \approx 0.23$, which is slightly less than what resulting from the asymptotic computation by Hilhorst (2005), who obtained a linear coefficient of 0.25. Secondly, and, more interestingly, we confirm that Desch's law is violated, *i.e.* $\langle \mu(P)|_\alpha \rangle_n \neq b_1(n + b_2)$, as shown, e.g. by Zhou (2001). Nevertheless, instead of a polynomial dependence on *n*, we find that a square root aw can be established, i.e. $\langle \mu(P)|_\alpha \rangle_n \approx c_1/\sqrt{\rho_0} (\sqrt{n + c_2})$. Our data give $c_1 \approx 1.71$, again slightly less than the asymptotic computation by Hilhorst (2005), who obtained $c_1 = \sqrt{\pi} \approx 1.77$. Fig. 8 features a log-log plot to emphasize such result. We note that the Lewis law and such law allow the establishment of a weakly *n*-dependent relationship such as $\langle \mu(A)|_\alpha \rangle_n \propto [\langle \mu(P)|_\alpha \rangle_n]^2$, which is re-ensuring and self-consistent at least in terms of dimensional analysis. Moreover, this agrees with the asymptotic result for large *n*, $\langle \mu(A)|_\alpha \rangle_n = \frac{1}{4\pi} [\langle \mu(P)|_\alpha \rangle_n]^2$, which descends from the fact that, as shown in Hilhorst (2005), cells tends to a circular shape.

In the intermediate range ($0.5 < \alpha < 2$), we have that the Lewis law and the square root law are not verified, and, quite naturally, the functions $\langle \mu(A)|_\alpha \rangle_n$ and $\langle \mu(P)|_\alpha \rangle_n$ get more and more similar to their Poisson-Voronoi counterparts as $\alpha$ increases.

A very interesting result is that for all values of $\alpha$, $\langle \mu(A)|_\alpha \rangle_{n=6} = \langle \mu(A)|_\alpha \rangle = \langle \mu(A)|_V \rangle$ (which implies that $a_1(6 + a_2) = 1 \Rightarrow a_2 = \frac{1}{a_1} - 6 \approx -1.65$) and $\langle \mu(P)|_\alpha \rangle_{n=6} = \langle \mu(P)|_V \rangle$ (which implies that $c_1\sqrt{6 + c_2} = 4 \Rightarrow c_2 = \left(\frac{4}{c_1}\right)^2 - 6 \approx -0.49$), whereas the ensemble mean estimators restricted to the other polygons are biased (positive bias for *n*>6 and negative bias for *n*<6). The fact that, in general, the statistics performed only on the most probable state coincides, at least in terms of ensemble mean, with that of the complete set of states, mirrors somehow the equivalence between the canonical and microcanonical formulations of the thermodynamics of a system. In this sense,

the number of sides seem to have the status of a *thermodynamic state variable* fluctuating about *n=6*, and, hexagons reinforce their status as being the *generic* polygon in a quite general family of Voronoi tessellations.

## Summary and Conclusions

This numerical study wishes to bridge the properties of the regular square and honeycomb hexagonal Voronoi tessellations of the plane to those generating from Poisson point processes, thus analyzing in a common framework symmetry-break processes and the approach to uniformly random distributions. This is achieved by resorting to a simple parametric form of random perturbations driven by a Gaussian noise to the positions of the points around which the Voronoi tessellation is created. The standard deviation of the position of the points induced by the Gaussian noise is expressed as $|\varepsilon| = \alpha/\sqrt{\rho_0}$, where $\alpha$ is the control parameter, the intensity $\rho_0$ corresponds to having about $\rho_0$ points, and about the same number of Voronoi cells, inside the unit square, and $1/\sqrt{\rho_0}$ is the natural length scale. We consider as starting points the regular square and honeycomb hexagon tessellations with intensity $\rho_0$, and change the value of $\alpha$ from 0, where noise is absent, up to 5. In this way, the probability distribution of points is in all cases periodic. For each value of $\alpha$, we perform a set of simulations, in order to create an ensemble of points and of corresponding Voronoi tessellation in the unit square, and compute the statistical properties of *n*, *A*, and *P*, the number of sides, the area and the perimeter of the resulting cell, respectively. The main results we obtain can be listed as follows:

- The symmetry break induced by the introduction of noise destroys the square tessellation, which is structurally unstable: already for an infinitesimal amount of noise the most common turns out to be a hexagon, whereas the honeycomb hexagonal tessellation is very stable and all Voronoi cells are hexagon for finite noise up within a certain range of $\alpha$ ($\alpha < 0.12$). Interesting signatures of the symmetry break emerge from a linear relationship between the standard deviation of the perimeter and the area of the Voronoi cells and the parameter $\alpha$;

- Already for a moderate amount of Gaussian noise (say $\alpha > 0.5$), memory of the specific initial unperturbed state is lost, because the statistical properties of the two perturbed regular tessellations are indistinguishable;

- In the case of perturbed square tessellation, for a specific intensity of the noise determined by $\alpha = \alpha_m \approx 0.25$, it is possible to minimize the mean perimeter-to-area ratio of the

Voronoi cells, whereas by choosing $\alpha = \alpha_M \approx 0.75$ we obtain the maximum perimeter-to-area ratio for both perturbed tessellations;

- For large values of $\alpha$ (say $\alpha > 2$), quite expectedly, the statistical properties of the perturbed regular tessellations converge, both in terms of ensemble mean and fluctuations, to those of the Poisson Voronoi process with the same intensity, since the points generating the tessellations are practically randomly and uniformly distributed in the plane;

- For all values of $\alpha > 0$, the 2-parameter gamma distribution does a great job in fitting the distribution of sides, area, perimeters of the Voronoi cells, the only exceptions being the singular distributions obtained for *n* in the case of perturbed honeycomb tessellation for $\alpha < 0.12$;

- For all values of $\alpha > 0$ the ensemble mean of mean number of sides, area and perimeter (except the latter for $\alpha < 0.5$) of the cells are remarkably constant, and the most common polygons result to be hexagons, whereas the ensemble mean of the standard deviations of these quantities increase steadily, in agreement with the transition to a more extreme random nature of the tessellation;

- The geometrical properties of *n*-sided cells change with $\alpha$ until the Poisson-Voronoi limit is reached for $\alpha > 2$; in this limit the Desch law for perimeters is confirmed to be not valid and a square root dependence on *n*, which allows an easy link to the Lewis law for areas, is established;

- The ensemble mean of the cells area and perimeter restricted to the hexagonal cells coincides with the full ensemble mean; this might imply that the number of sides acts as a *thermodynamic state variable* fluctuating about *n=6*, and this reinforces the idea that hexagons, beyond their ubiquitous numerical prominence, can be taken as *generic* polygons in 2D Voronoi tessellations.

In previous works much larger densities of points have been considered – up to several million (Tanemura 2003). In this work, those numbers would be rather inconvenient because we perform a parametric study of ensemble runs. Nevertheless, we wish to emphasize that the choice of $\rho_0$ does not alter any of the result on the ensemble mean of statistical properties of *n*, *A*, and *P*. In fact, for all values of $\alpha$, and not only in the Poisson-Voronoi limit, as discussed in the paper and verified in several simulations to hold accurately for $\rho_0$ up to 1000000, $\langle \mu(A)|_\alpha \rangle, \langle \sigma(A)|_\alpha \rangle, \langle \mu(A)|_\alpha \rangle_n$ scale as $1/\rho_0$ and $\langle \mu(P)|_\alpha \rangle, \langle \sigma(P)|_\alpha \rangle, \langle \mu(P)|_\alpha \rangle_n$ scale as $1/\sqrt{\rho_0}$, whereas $\langle \mu(n)|_\alpha \rangle$ does not depend on $\rho_0$. Therefore, by multiplying these quantities times the appropriate power of $\rho_0$, we get universal

functions. Where, instead, the choice of $\rho_0$ is more relevant is in the pursuit for a small ratio between the ensemble fluctuations and the ensemble mean of the above mentioned quantities, because the ratio decreases with $\rho_0$, as to be expected. A related benefit of a larger value of $\rho_0$, is the possibility of computing the statistics on *n*-sided cells on a larger number of classes of polygons, since the probability of detecting a *n*-sided polygons decreases very quickly with *n*.

We believe that it is definitely worthy to extend this study to the 3D case, which might be especially significant for solid-state physics applications, with particular regard to crystals' defects and electronic impacts of vibrational motion in various discrete rotational symmetry classes. Nevertheless, a much larger computational cost has to be expected, since a larger number of points and a larger computing time per point are required for sticking to the same precision in the evaluation of the statistical properties.

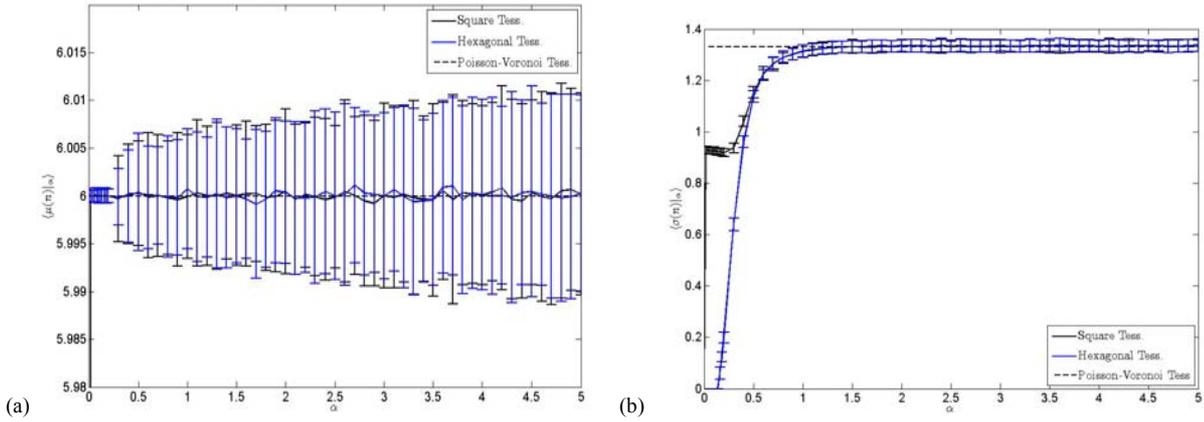

**Figure 1:** Ensemble mean of the mean - (a) - and of the standard deviation – (b) – of the number of sides (*n*) of the Voronoi cells. Note that in (a) the number of sides of all cells is 4 (out of scale) for α=0 in the case of regular square tessellation. Half-width of the error bars is twice the standard deviation computer over the ensemble. Poisson-Voronoi limit is indicated.

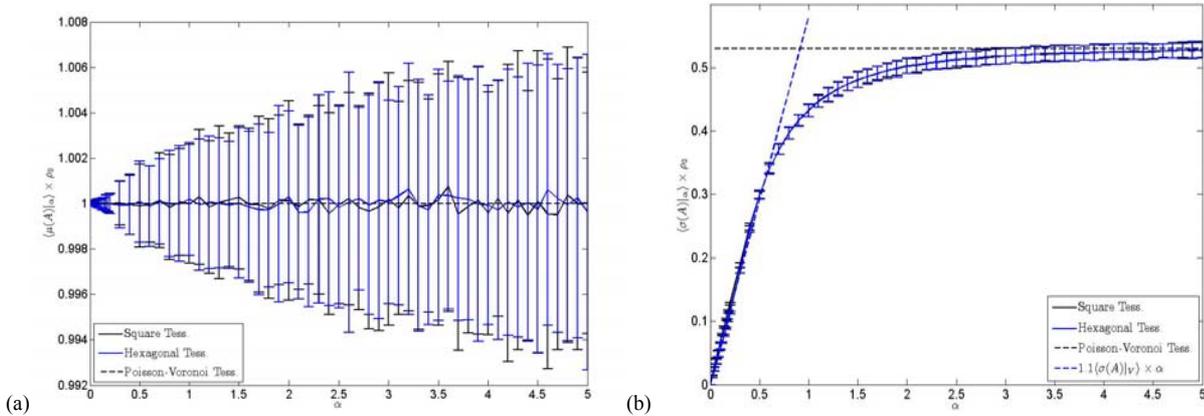

**Figure 2:** Ensemble mean of the mean - (a) - and of the standard deviation – (b) – of the area (*A*) of the Voronoi cells. Half-width of the error bars is twice the standard deviation computer over the ensemble. Poisson-Voronoi limit is indicated. In (b), linear approximation for small values of *α* is also shown. Values are multiplied times $\rho_0$ in order to give universality to the ensemble mean results.

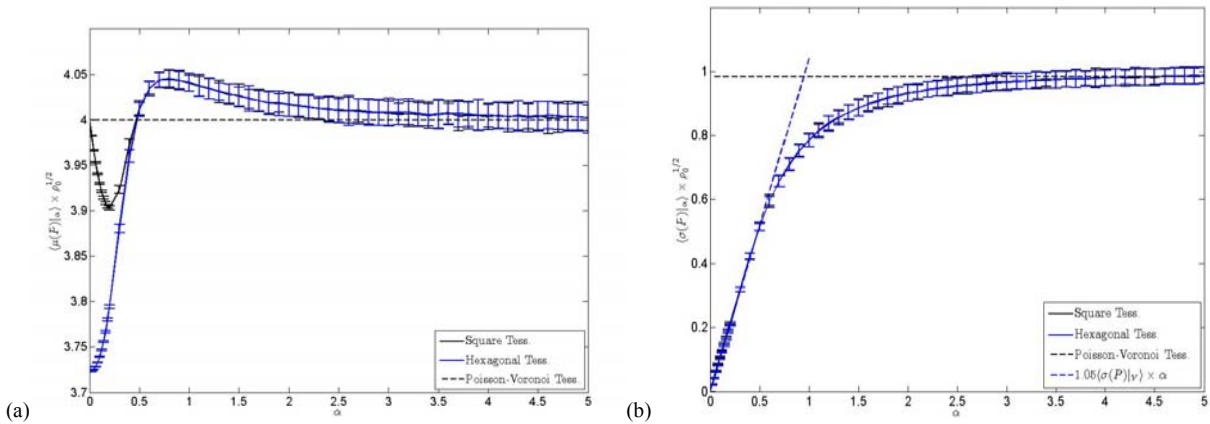

**Figure 3:** Ensemble mean of the mean - (a) - and of the standard deviation – (b) – of the perimeter (*P*) of the Voronoi cells. Half-width of the error bars is twice the standard deviation computer over the ensemble. Poisson-Voronoi limit is indicated. In (b), linear approximation for small values of *α* is also shown. Values are multiplied times $\rho_0^{1/2}$ in order to give universality to the ensemble mean results.

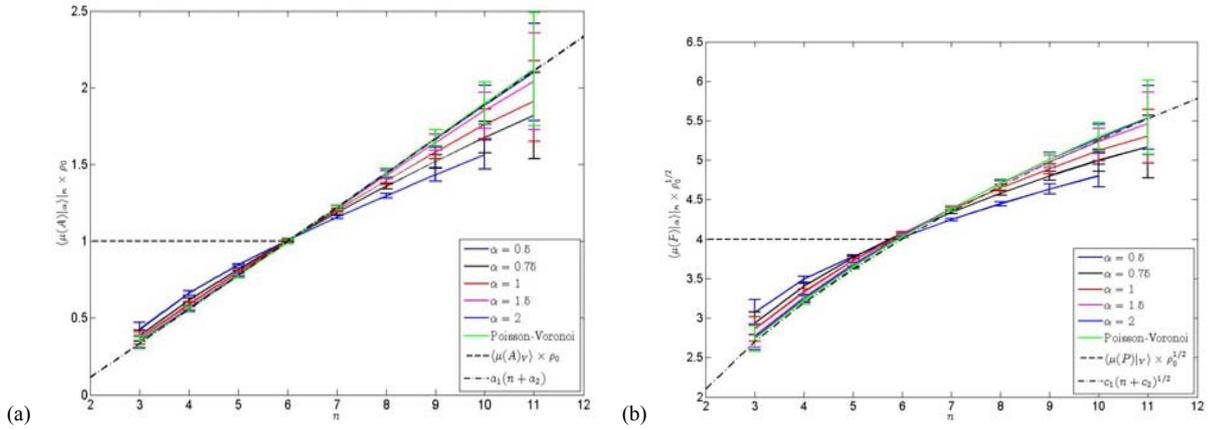

**Figure 4:** Ensemble mean of the area $A$ - (a) - and of the perimeter $P$ – (b) – of $n$-sided Voronoi cells. Half-width of the error bars is twice the standard deviation computer over the ensemble. Full ensemble mean is indicated. Linear (a) and square root (b) fits of the Poisson-Voronoi limit results as a function of $n$ is shown. Values are multiplied times $\rho_0$ (a) and $\rho_0^{1/2}$ (b) in order to give universality to the ensemble mean results.